# Myriad of Terahertz Magnons with All-Optical Magnetoelectric Functionality for Efficient Spin-Wave Computing in Honeycomb Magnet $Co_4Ta_2O_9$

*Brijesh Singh Mehra[1], Sanjeev Kumar[1], Gaurav Dubey[1], Ayyappan Shyam[1], Ankit Kumar[1], K Anirudh[1], Kiran Singh[2], Dhanvir Singh Rana[1]\**

[1]Indian Institute of Science Education and Research Bhopal, 462066, India

[2]Dr. B. R. Ambedkar National Institute of Technology, Jalandhar, 144011, India

Email: dsrana@iiserb.ac.in

## Abstract

Terahertz (THz) magnonics represent the notion of mathematical algebraic operations of magnons such as addition and subtraction in THz regime – an emergent dissipation-less ultrafast alternative to existing data processing technologies. Spin-waves on antiferromagnets with a twist in spin order host such magnons in THz regime, which possess advantage of higher processing speeds, additional polarization degree of freedom and longer propagation lengths compared to that of gigahertz magnons in ferromagnets. While interaction among THz magnons is the crux of algebra operations, it requires magnetic orders with closely spaced magnon modes for easier experimental realization of their interactions. Herein, rich wealth of magnons spanning a narrow energy range of 0.4-10 meV is unraveled in $Co_4Ta_2O_9$ using magneto-THz spectroscopy. Rare multitude of ten excitation modes, either of magnons or hybrid magnon-phonon modes is presented. Among other attributes, spin-lattice interaction suggests a correlation among spin and local lattice distortion, magnetostriction, and magnetic exchange interaction signifying a THz magnetoelectric effect. This unification of structural, magnetic and dielectric facets, and their magnetic-field control in a narrow spectrum unwinds the mechanism underneath the system's complexity while the manifestation of multitude of spin excitation modes is a potential source to design multiple channels in spin-wave computing based devices.

# 1. Introduction

The collective precessions of spins in a magnetically ordered material, known as spin waves with quanta as magnons, is fundamental to the pursuit of next generation low dissipation and ultrafast device operation. The non-ohmic propagation of spin waves and terahertz (THz) frequency control of antiferromagnetic (AFM) spins are the cornerstones for areas of magnonics and THz spintronics. The current focus is on the control and manipulation of the spin-waves/magnons for information processing with their use in logic-devices, low-resistance circuits, ultrafast computing, and so on.[1,2] The magnons also play a prominent role in underlying magnetic symmetric and/or antisymmetric exchange interaction and magnetic anisotropy and derive exotic non-trivial magnetic and quantum phases such as topological phases, quantum spin liquid, etc.[3,4] As this rich magnon physics in condensed matter systems is envisaged to boost the information processing technology, the search is on for materials with robust magnons along with reliable means of their control and propagation. Magnetoelectric materials (ME) could provide such platform in a way that the non-collinear magnetic order induced electric/dielectric phase will facilitate the electric and magnetic field tunable magnons in the THz frequency range.

A popular ME system possessing non-collinear magnetic order responsible for its mutual control of magnetic and electric orders, $Co_4Ta_2O_9$ crystallize in α-$Al_2O_3$ type trigonal structure (space group $P\bar{3}c1$) with Co and Ta occupation ratio 2:1.[5] Here, $Co_4Ta_2O_9$ exhibits ME phase below Neel's temperature ($T_N$ ~20 K), wherein Co occupies two inequivalent sites [Co(I) and Co(II)] responsible for its magnetic order. Investigations using Neutron diffraction has established that $Co^{2+}$ spins lie in the basal plane contrary (magnetic space group C2/c') to the previous notion of the spins aligning along the trigonal axis.[5,6] Another investigation employing a combination of neutron diffraction and directional magnetic susceptibility reassigned the magnetic space group in $Co_4Ta_2O_9$ to be C2'/c.[7] It exhibits diverse properties such as i) dielectric anomaly at $T_N$ and its enhancement with applied magnetic field,[8] ii) shearing mode of cobalt ions which couples via interlayer interaction,[9] iii) complex magnetic state (weakly ferromagnetic or/and glassy state) below 10 K,[7,10] iv) magnetic field induced electric polarization,[8] v) nonlinear ME effect above spin-flop transition for in-plane magnetic fields,[10] *etc*. All these myriads of complex structural, dielectric, and magnetic properties and intercorrelation amongst them are expected to host a variety of spin-excitations due to the non-collinear nature of its AFM order. However, any experimental demonstrations of spin wave/magnons around the 'Γ' point either by inelastic neutron scattering or THz spectroscopy are yet to be made. Insights on spin excitations shed light on the detailed complexity of exchange interactions that stabilizes the magnetism.

The low-energy attribute of THz radiation makes it uniquely sensitive to probe electric and magnetic phases. This combined with its spectral range appropriate to host spin-excitation modes makes it a powerful tool to investigate the presence of spin waves as well as the dynamics of electric/dielectric medium underneath. This versatile contactless technique spans not only the energy range of variety of quasiparticles in condensed matter system such as low-lying phonon mode, charge density waves, Higg's mode, superconducting gap and so on[11–14] but also probes various symmetric and anti-symmetric magnetic interactions which are the building blocks of magnetic Hamiltonian in physical sciences. Here, we report a record *ten* excitations pertaining to magnon, phonon, and hybridized magnon-phonon modes in $Co_4Ta_2O_9$ using magneto-THz-time domain spectroscopy [Figure 1]. Zero-field ME ground state unveiled rich wealth of spin-excitations including multiple gapped modes and a pure lattice vibration which couples with the magnetic structure at the AFM transition temperature. We demonstrated that magnetic ions' lattice displacement is vital in accounting for magnetically induced polarization. Theoretical spin-wave calculations were performed to determine the strength of exchange interactions. Evidence of optical ME effect are also presented along with the multitude of magnon and magnon-phonon modes.

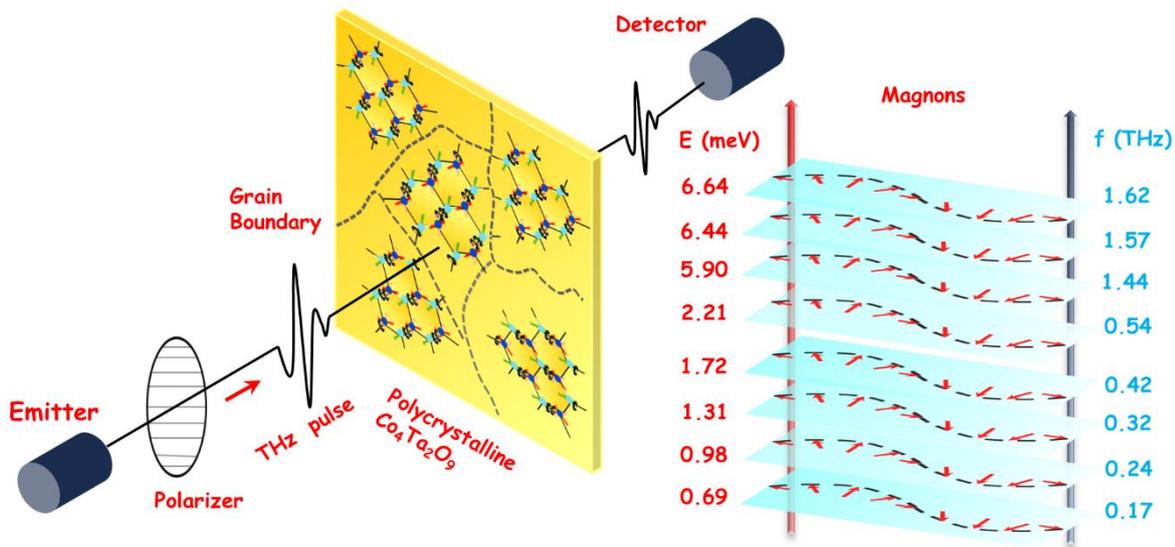

**Figure 1:** Schematic depiction of experimental set up of magneto-THz time-domain spectroscopy and the observation of nine temperature-dependent magnons in the polycrystalline sample of $Co_4Ta_2O_9$.

## 2. Results and Discussions

The susceptibility versus temperature data shows a Neel's temperature ($T_N$) of 20.5 K and a complex magnetic transition at 10 K which corroborates well with the previous reports[8,15,16] [Figure 2(a) and Inset Figure 2(a)]. THz response in three different magnetic phases [Figure 2(b)], namely, 20.5 K in paramagnetic region, 13 K in AFM region, and 6 K in complex magnetic region displays distinct features. The strength of periodic THz oscillation in higher time scale (64-74 ps) shows abrupt increase at the onset of these regions. The time scale of this periodicity is approximately 2 ps in AFM state. This feature is absent above $T_N$.

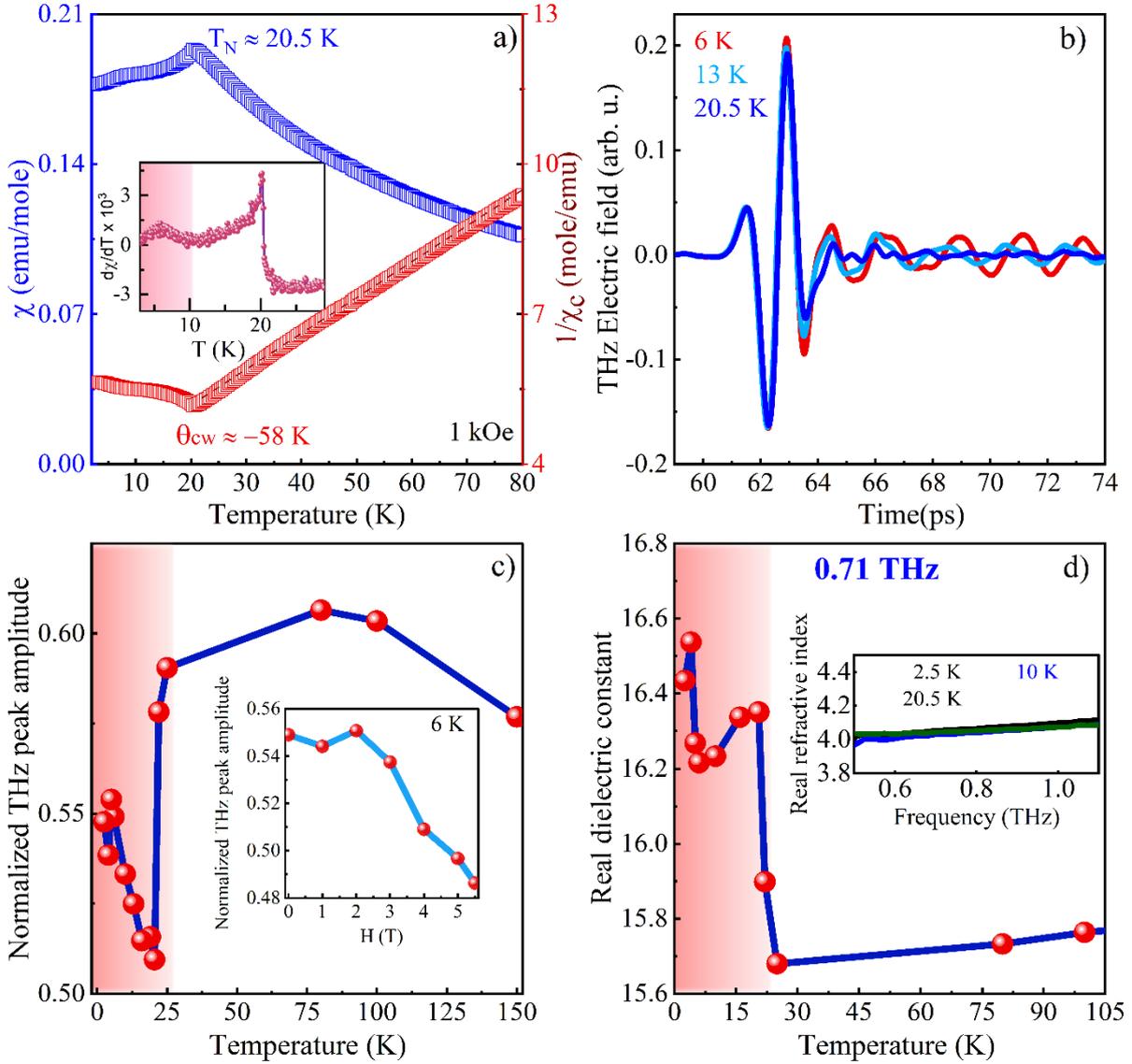

**Figure 2:** a) Magnetic susceptibility versus temperature. b) THz electric field at three different temperatures. c) Normalized THz peak amplitude as a function of temperature and magnetic field (Inset), respectively. d) Real dielectric constant versus temperature at 0.71 THz. Inset shows refractive index in the THz frequency range.

The normalized THz peak amplitude defined as the ratio of THz electric field peak position with and without the sample, $\frac{E_{(sample)peak}(t,T)}{E_{(reference)peak}(t,T)}$, displays a sudden drop of ~12 % at the $T_N$ [Figure 2(c)] depicting a sensitivity of THz electric field to the spin-order at the magnetic transition. This feature combined with a large magnetic field dependence of normalized THz electric-field peak at 6 K [Inset Figure 2(c)] are unambiguous evidence of THz ME effect in this system. Also, The THz data yields a real refractive index of ~ 4 [Inset Figure 2(d)], which agrees well with the literature.[8] The real dielectric constant [at $\omega$ = 0.71 THz] increases with decreasing temperature and exhibits anomalies at both the magnetic transitions at 20.5 and 10 K [Figure 2(d)], which is consistent with the behavior of magnetization data [Inset Figure 2(a)].

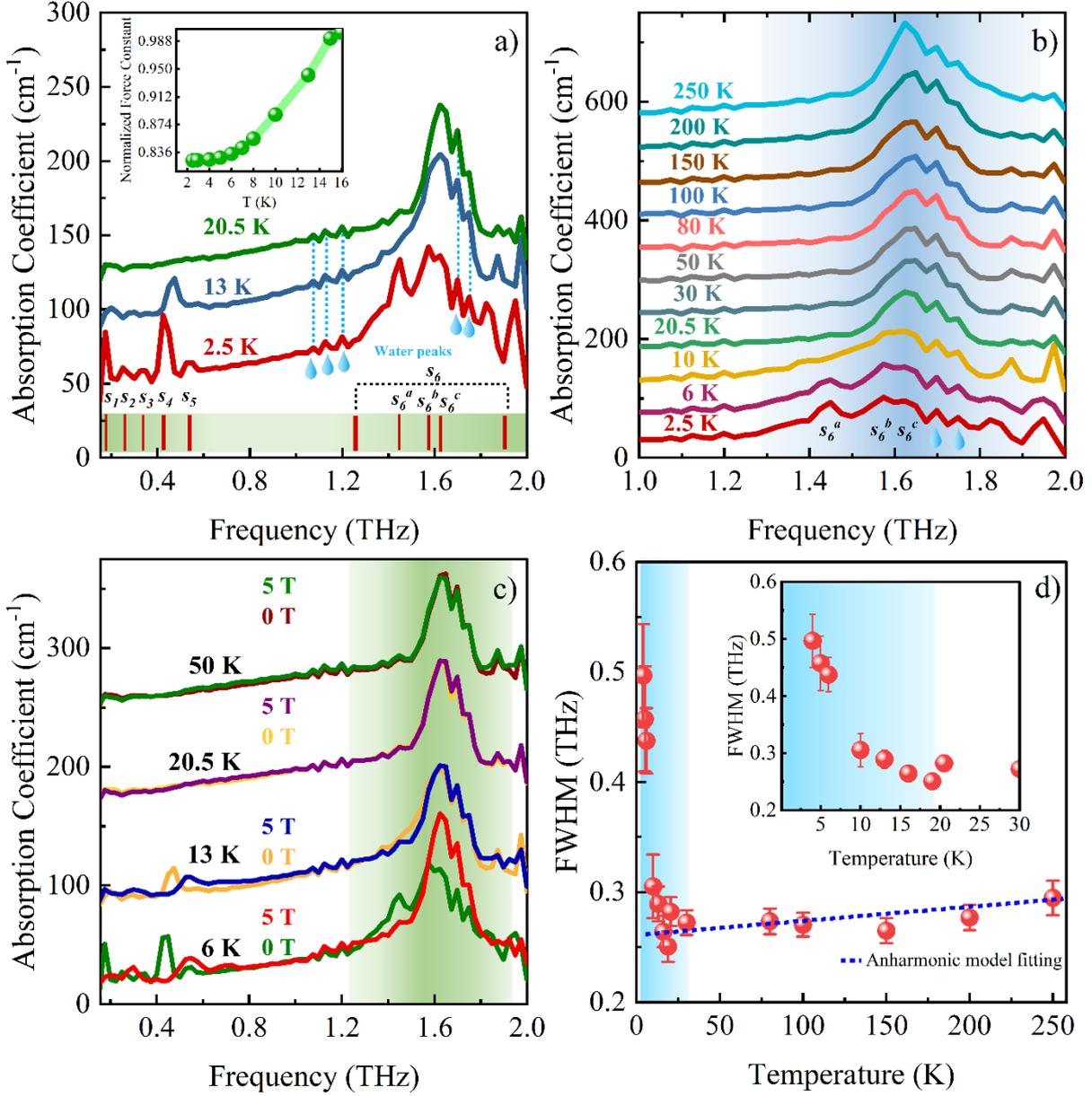

**Figure 3:** a) Absorption coefficient versus THz frequency. Normalized force constant ($k_N=k_T/k_{16K}$; where $k_T$ is the force constant at temperature T and $k_{16\,K}$ is the force constant at 16 K) as a function of temperature is depicted in inset. b) Temperature dynamics of $s_6$ mode. c) Absorption coefficient versus THz frequency at 6, 13, 20.5, and 50 K with magnetic field 0 and 5 T. d) FWHM of $s_6$ mode as a function of temperature. Inset shows the variation in temperature range 2-30 K. For clarity offset is provided in a, b, and c.

In $Co_4Ta_2O_9$, the magnetic symmetry lowers from trigonal in the paramagnetic state to monoclinic symmetry in the AFM state. Due to a large in-plane anisotropy, spins lie in the basal plane with an in-plane canting angle of 14° between Co(I) and Co(II) ions.[7,15] Figure 3(a) depicts the temperature-dependent THz absorption spectra of $Co_4Ta_2O_9$ in the frequency range 0.1-2.1 THz. As is evident, a large number of resonance absorption peaks, expectedly spin wave excitations/magnons, manifest in two different regimes of the spectra. Below $T_N$, THz absorption spectra reveals three excitations, namely, *s4*, *s5*, and *s6* (broad mode) at 0.42, 0.54, and 1.2-1.9 THz, respectively. Below 10 K, additional excitations *s1*, *s2*, *s3*, *s6$^a$*, *s6$^b$*, and *s6$^c$* emerge sharply at 0.17, 0.24, 0.32, 1.44, 1.57, 1.62 THz, respectively. It may be seen that these sharp excitations (*s6$^a$*, *s6$^b$*, and *s6$^c$*) are superimposed on *s6* mode. These additional modes in THz spectra are in accordance with the distinct wiggles in the THz electric field tail (64 to 74 ps) at 13 and 6 K corresponding to these resonances [Figure 2(b)]. Accounting all these sharp and broad modes, it may be noted that the zero-field ground state of $Co_4Ta_2O_9$ at 2.5 K exhibits **nine excitations** in a narrow frequency range of 0.1-2.2 THz which is significantly larger than three zero-field excitations in $Co_4Nb_2O_9$.[17] This rare manifestation of closely spaced multitude magnetic excitations in the THz frequency is potentially relevant for fundamental and applied pursuits in the field of antiferromagnetic magnonics algebra.[32-33]

Now we shed light on the origin and detailed scrutiny of thus observed magnetic modes. Starting with broad mode *s6*, we observe that it continues to grow beyond magnetic ordering [Figure 3(b)]. In the paramagnetic region, the red shift in this peak position and a larger full-width half-maximum (FWHM) [Figure 3(d)] with increasing temperature are its attributes that point towards the pure phonon mode. This was confirmed in the paramagnetic state, [Figure 3(c)] where the applied field does not affect the structure and strength of this mode. In contrast, it shows profound field-induced changes in the magnetic ordered region below $T_N$ which points towards coupling of the lattice and spin waves, consequently, giving rise to magnon-phonon excitation. This is further evidenced by the deviation of FWHM from the cubic anharmonicity of phonon-linewidth [Figure 3(d)] (in accordance with phonon-phonon anharmonic model) defined as[18]

$$\Gamma = \Gamma_0 + C(1 + \frac{2}{e^{\hbar\omega_0/2k_BT}-1}) \qquad (1)$$

where, $\omega_0$ and $\Gamma_0$ are the mode frequency and linewidth at absolute zero temperature, respectively. This demonstration of low-lying phonon mode and its entanglement with spins via magnetic field is an important lead to unravel its ME character in a later section.

Now we turn our attention to multiple gapped excitations *s4* and *s5* which we assign to pure spin-excitation / magnon modes, for the following reasons. As the magnetic symmetry lowers, below $T_N$, the gapped modes emerge with their origin expected in strong in-plane single-ion anisotropy. To verify this, we collected the THz spectra of $Mn_4Ta_2O_9$, which is isostructural equivalent of $Co_4Ta_2O_9$. Unlike $Co_4Ta_2O_9$, $Mn^{2+}$ spins lie along the trigonal axis in the AFM state.[19,20] Its THz spectra clearly shows the absence of '*s4* and *s5*' type gapped modes at 10 K [Figure SI 4; Supporting Information], elucidating that the origin of gapped mode to be associated with the basal in-plane single-ion anisotropy of $Co_4Ta_2O_9$. Normalized force constant as a function of temperature unveils the softening of s4 mode as temperature is lowered [Inset Figure 3(a)].

The magnon and phonon lifetime is an important factor in contemplating various THz magnonics based devices. In the present case, this was calculated using energy-time uncertainty principle.[21] For the phonon excitation, its linewidth decreases with decreasing temperature below $T_N$ owing to reduced strength of thermal fluctuation, phonon-phonon scattering, and anharmonic effect [Figure 3(d)]. Phonon lifetime [Figure SI 5(a)] in magnetic ordered phase (1 ± 0.09 ps) is smaller than that in paramagnetic phase (1.83 ± 0.07 ps). This is because at $T_N$ the linewidth begins to deviate from the pure phonon vibration as this phonon couples with magnons and hence suffers additional scattering mechanism. For the gapped spin-excitation, the strength of relevant mode should strengthen as the temperature is lowered. Exact trend can be observed in Figure SI 5(b), where the spin-gapped mode (*s4*) lifetime increases as temperature is lowered implying a long coherent length in the low temperature regime with a lifetime of 17.21 ± 2.99 ps at 2.5 K.

Using the THz gapped mode, *s4*, we tried to estimate the magnetic exchange interaction and magnetic anisotropy for $Co_4Ta_2O_9$. The gapped mode originates from single ion anisotropy, $E_{gapped} = 4S\sqrt{JD}$, where D, J, S are the single-ion anisotropy constant, nearest neighbor interaction, and spin moments, respectively.[22] We assume that the nearest neighbor interaction in isostructural $Co_4Nb_2O_9$ and $Co_4Ta_2O_9$ are same ($J_{Co4Nb2O9}$ = -0.7 meV and $D_{Co4Nb2O9}$ = 1.8 meV).[23] From THz experiments, $(E_{gapped})_{Co4Nb2O9}$ = 3.15 meV [Supporting Information] and, $(E_{gapped})_{Co4Ta2O9}$ = 1.72 meV. Using the above relation, the single-ion anisotropy constant ($D_{Co4Ta2O9}$) was calculated to be 0.53 meV, which turns out to be less than the absolute value of nearest neighbor exchange interaction $J_{Co4Ta2O9}$ = 0.7 meV. However, it is inadmissible for the following reasons: i) $Co_4B_2O_9$ (B=Nb, Ta) possesses large in-plane anisotropy such that even large value of external magnetic field along c direction cannot flop the spin from basal plane, which implies that D>J[6,23,24] and, ii) as per Goodenough-Kanamori-Anderson (GKA) rules, the super exchange interaction J is proportional to $t^2/U$, where t is effective orbital hopping and U is the Hubbard repulsion.[25,26] The first-principle studies[15] suggest that $U_{Co4Nb2O9} < U_{Co4Ta2O9}$ and Co are more localized in $Co_4Ta_2O_9$, which implies a small spatial extent of the electron wavefunction and a reduced overlap between adjacent atomic orbital; hence, $J_{Co4Ta2O9} < J_{Co4Nb2O9}$. Therefore, our assumption in similarity of nearest neighbor interaction does not hold suggesting that these systems behave differently. It is required that $D_{Co4Ta2O9} > J_{Co4Ta2O9}$ and $J_{Co4Ta2O9} < J_{Co4Nb2O9}$.

To determine the magnetic exchange interactions, a spin Hamiltonian was formulated for the honeycomb magnet $Co_4Ta_2O_9$ which can be written as[23]

$$H = H_c + H_{p,b,pb} + \sum_{Co(I)} D\,(S_{iz}^I)^2 + \sum_{Co(II)} D\,(S_{iz}^{II})^2 + \sum_{<i,j>} D_{ij} \cdot (S_i \times S_j) \qquad (2)$$

where $H_c$ is the Heisenberg term for the nearest and next nearest neighbor along c; $H_{p,b,pb}$ are the Heisenberg terms for the planar, buckled, and planar-buckled networks. Third and fourth term represent single-ion in plane anisotropy where I and II denote two inequivalent sites of cobalt [Co(I) and Co(II) sites] and $D_{ij}$ represent the DM interaction.

Here, the magnetic exchange interaction, single-ion anisotropy strength and high-resolution observations of magnons at Γ point by THz spectroscopy underpins the magnetic structure and the magnonic dynamics. To get insights of the Brillouin zone, beyond the Γ point, spin-wave calculations were performed using the above spin Hamiltonian in SPINW.[27] In-plane magnetic structure was considered as revealed by neutron diffraction experiments [28] and using the THz Γ point data, the magnetic exchange interactions were computed. The average powder spin-wave spectra for $Co_4Nb_2O_9$ [Figure 4(a)] shows gapped and gapless excitations which matches accurately with the Γ point of THz data. The magnetic exchange interactions ($J_{Co4Nb2O9}$=-0.6 meV and $D_{Co4Nb2O9}$=1.6 meV) too agree well with those depicted by the inelastic neutron experimental results.[28] Using magnetic structure of temperature above 10 K for $Co_4Ta_2O_9$[29], the powder averaged spin-wave calculations yielded gapped excitations and magnetic exchange interactions $J_{Co4Ta2O9}$=-0.4 meV and $D_{Co4Ta2O9}$=1.1 meV. These are at lower energy and weaker, respectively, as compared to those of $Co_4Nb_2O_9$ and in perfect agreement with the condition $D_{Co4Ta2O9} > J_{Co4Ta2O9}$ and $J_{Co4Ta2O9} < J_{Co4Nb2O9}$ [Figure 4(b)].

In contrast to above-mentioned low-energy spin-excitations, the higher energy modes (> 5 meV) obtained from spin-wave calculations and THz experimental data do not agree for either of $Co_4Ta_2O_9$ and $Co_4Nb_2O_9$. This is because the spin Hamiltonian accounts only for spin-excitations while the phonon contribution as well as spin-lattice coupling terms are not incorporated. This deviation is further verification of broad mode $s_6$ being [Figure 3(a)] a spin-phonon coupled mode. As Nb is systematically replaced by Ta in $Co_4Ta_2O_9$ [Figure 4(c)], the gapped mode (*s4*) gradually shifts to higher energies owing to gradual enhancement of the magnetic exchange interactions (J and D) [Figure 4(d)]. Thus, magnetic exchange interactions play a prominent role in driving the spin-excitations.

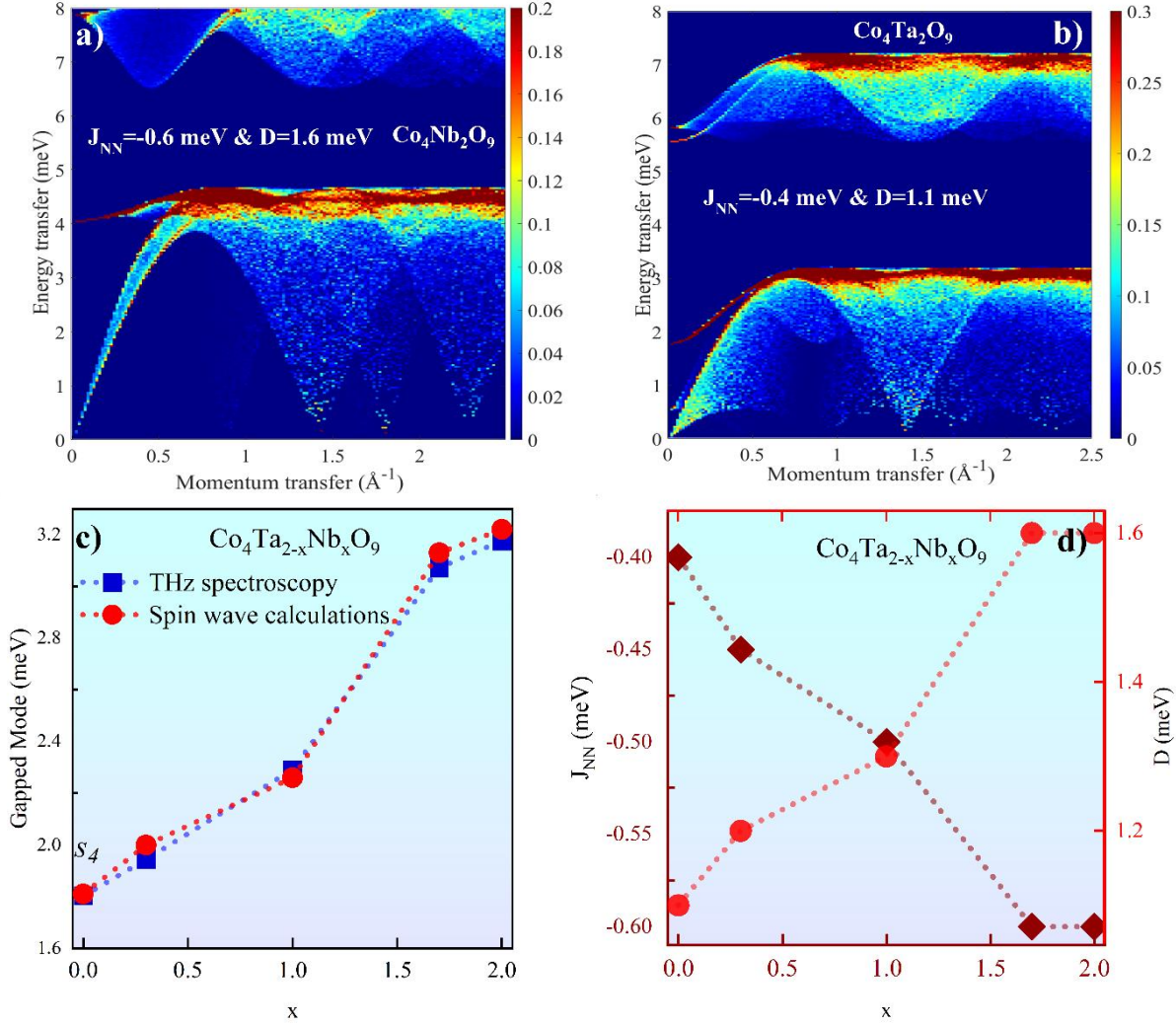

**Figure 4:** Spin wave calculations for a) $Co_4Nb_2O_9$ and b) $Co_4Ta_2O_9$. c) Comparison of experimental and calculated value of Gapped mode *s4*, at the $\Gamma$ point, as a function of Nb doping in $Co_4Ta_{2-x}Nb_xO_9$. d) Obtained value of magnetic exchange interactions (J=nearest neighbor interaction and D=single-ion anisotropy) for $Co_4Ta_{2-x}Nb_xO_9$ (x=0, 0.3, 1, 1.7, 2).

Now, to shed light on the origin of excitations *s1*, *s2*, *s3*, *s6$^a$*, *s6$^b$*, and *s6$^c$* (Figure 3a), it is imperative to invoke comparison of some relevant properties of $Co_4Ta_2O_9$ from $Co_4Nb_2O_9$. As noted in previous section, weaker exchange interactions in $Co_4Ta_2O_9$ compared to that in $Co_4Nb_2O_9$ render it a softer magnet. In the latter, there has been no indication of structural transition even at 5 K, [6,7,16,24] though the possibility of local lattice distortion and magnetostriction was never ruled out.[15] In the case of $Co_4Ta_2O_9$, however, the structural transitions, larger local lattice distortions and magnetostriction becomes more promising owing to its sensitivity to change in magnetic structure due to weaker magnetic interactions. This deviation of $Co_4Ta_2O_9$ from $Co_4Nb_2O_9$, below 10 K, is the source of non-linear ME response[30] and complex magnetic state[29,30] (in $\chi$-T curve) in the former. However, its magnetic structure is reported only down to a minimum temperature of 15 K,[29] which is higher than the complexity-rich magnetism regime of below 10K. It is this low temperature regime, wherein the complex magnetic state, magnetostriction and local lattice distortion in $Co_4Ta_2O_9$ create entangled magnetic moments and lattice vibrations, which consequently give rise to six spin excitations *s1*, *s2*, *s3*, *s6$^a$*, *s6$^b$*, and *s6$^c$* in the THz spectra [Figure 3(a, b)]. No such excitations have been observed in $Co_4Nb_2O_9$. The $Co_4Ta_2O_9$, thus, hosts a ground for unique correlation between magnetic exchange interactions, local lattice distortion, and magnetostriction phenomena.

As $Co_4Ta_2O_9$ is a magnetic-field induced ME system, it is imperative to understand the magnetic-field control of THz spin-wave excitation and the ME character. Magnetic field dependence of THz spectra is plotted in Figure 5 (a). Spin-excitations *s1*, *s2*, *s3*, *s6$^a$*, *s6$^b$* and *s6$^c$* get suppressed with increasing magnetic field which reflects the expected magnetically malleable structure of $Co_4Ta_2O_9$. At 2T, all these excitations annihilate while *s4* and *s5* shift to higher frequencies. Derived from the gapped mode (*s4*), the normalized force constant ($k_H/k_{0T}$; $k_H$ is the force constant at magnetic-field H and $k_{0T}$ is force constant at zero magnetic-field) indicates the hardening of the magnetic coupling between neighboring spins as a function of increasing magnetic-field [Inset Figure 5(a)]. As magnetic field exceeds 3 T, mode *s7* appears in the detectable range of our THz spectra. This excitation, identified as goldstone (/gapless) mode, appears in the microwave region due to spontaneous symmetry-breaking at $T_N$ and it shifts linearly towards higher THz frequency with increasing field [Figure 5(a) inset]. This gapless mode is shown in the spin-excitation simulations as well as in the THz spectra. The behaviour of peak frequency of this gapless excitation yields Landé g-factor g=3.09 for $Co_4Ta_2O_9$ and g=2.68 for $Co_4Nb_2O_9$ [Figure SI 3 and 7, Supporting Information]. Clearly the value of Landé g-factor suggests unquenched orbital moments both in $Co_4Ta_2O_9$ and Nb counterpart.

The mechanism of the field-induced polarization in this series has been of great interest to understand the induced ME character. Knowledge of spin-phonon mode and its magnetic-field dependence from THz spectroscopy can provide valuable insights into this process. At 6 K, the peak strength of spin-phonon mode of $Co_4Ta_2O_9$ increases with an increase in magnetic field [Figure 5 (a, b)] which is associated with the field induced electric polarization. As lattice distortion, magnetostriction, and spin-phonon coupling are highly inter-related in these materials, the induced electric polarization can be explained as follows. At zero magnetic-field, below $T_N$, the presence of spin-phonon coupling suggests the entanglement of spin and lattice. However, the Co ions hold two inequivalent sites Co(I) and Co(II) which are in centrosymmetric positions with respect to trigonal axis, hence, lacks any net polarization. However, on the application of magnetic-field the strength of spin-phonon coupling increases [Figure 5 (b)]. This expectedly displaces the magnetic ions from their centrosymmetric positions, resulting in the manifestation of magnetically induced polarization. This scenario of field-induced displacement of Co-ion is depicted in Figure 5 (c) As the mass of Ta is larger than Nb (ions to which Co-ions are bonded), the effective displacement in case of $Co_4Ta_2O_9$ is less than that in $Co_4Nb_2O_9$. This also explains why magnetic field induced polarization in $Co_4Nb_2O_9$ is more as compared to $Co_4Ta_2O_9$. The THz characteristic features of the magnetic resonances too possess this mass effect where with increasing concentration of Nb the resonances are shifting to higher THz frequency [Figure SI 8]. To surmise, this entire mechanism based on modulation of magnetic exchange interactions provides a pathway to control the energy of spin-gapped modes, spin-phonon coupled mode, and phonon mode whereas the effective lattice displacement of cobalt ions is responsible for magnetic-field induced polarization.

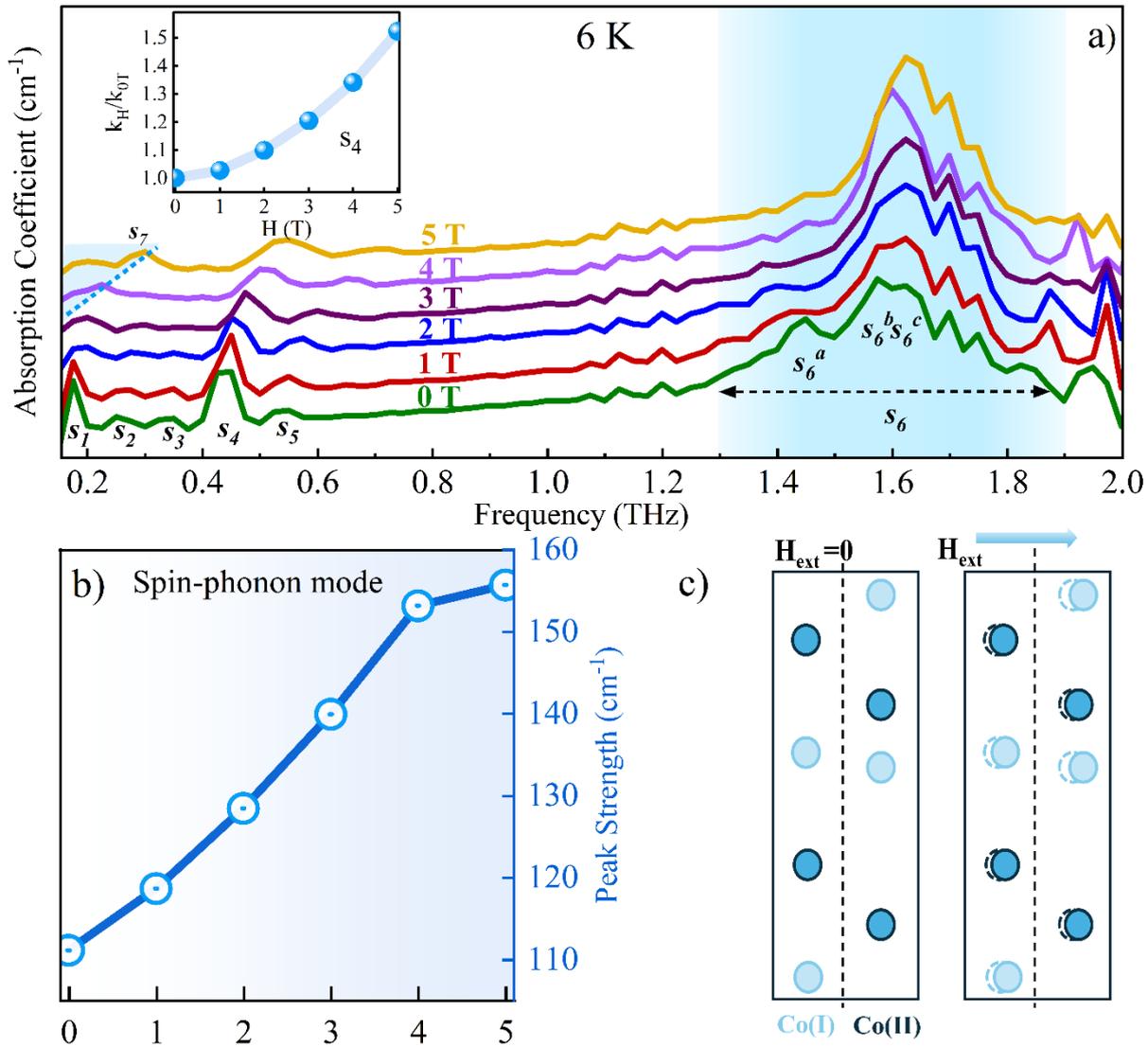

**Figure 5:** a) Absorption coefficient versus THz frequency at 6 K with varying magnetic field (Offset is provided for clarity). Inset highlights the normalized force constant ($k_H/k_{0T}$; $k_H$ is force constant at magnetic-field H and $k_{0T}$ is force constant at zero magnetic-field) derived from the peak position of $s_4$ gapped mode. b) Peak strength of spin-phonon mode as a function of magnetic field and c) schematic illustrating magnetic-field induced polarization.

From the applied pursuit, the wave nature of magnons (spin-waves) offers a pathway to encode information in amplitude, phase, or the combination of both [Figure 6(a)] which is at core of non-Boolean algebra driven spin-wave computation.[31-33] The spin excitations in THz regime offer two novel characteristics, namely, the THz magnons propagate with ultra-low dissipation as it does not involve flow of electric charge, and with ultrafast speeds, both of which are much desired attributes for futuristic technologies. The existence of multiple magnon modes, as demonstrated in this work, is a pre-requisite for efficient data transfer spin-wave logic operations. In another facet, in the direction of spin-wave computation the multitude of spin waves in $Co_4Ta_2O_9$ present itself as a potential candidate for multifrequency channeling in a narrow bandwidth of 0.1 – 2.5 THz. Such multi-channels allow for simultaneous transmission of multiple signals at different frequencies, increasing the overall data capacity. A schematic serving as a proof-of-concept for terahertz-magnonics-electronics multifrequency channeling is shown in Figure 6(b). Here, the frequency of THz radiation drives the resonant condition corresponding to that magnon which carries the information and provides the output as electronic signal via a spin wave to charge converter. Experimental realization of this concept requires systems having multiple magnons or other hybrid modes in THz regime. In this work, the magnetoelectric systems with non-collinear magnetic order prone to strong spin-lattice interactions provide appropriate platform for THz magnonics.

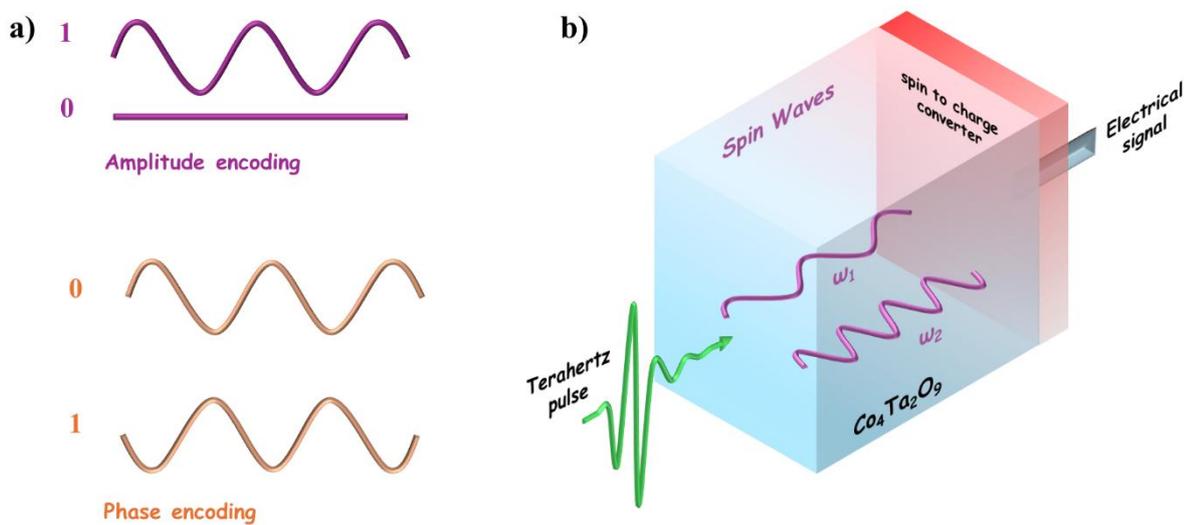

**Figure 6:** a) Types of information encoding via spin-waves: Amplitude and Phase encoding. b) Proof-of-concept of terahertz-magnonic-electronics multifrequency channelling device.

## 3. Conclusions

A myriad of low-energy excitations in $Co_4Ta_2O_9$ probed using magneto-THz spectroscopy evidence a remarkable host magnetoelectric system with a rare multitude of ten excitations comprising of magnon, phonon, and hybridized magnon-phonon modes. The THz probes and non-collinear magnetism further combine to unravel a THz magnetoelectric effect; a novel functionality not known to manifest at such high frequencies so far. The origin of magnon in a strong basal-plane anisotropy emphasized the structural, magnetic and electronic controls to all excitation modes. These experimental data are supported by theoretical spin-wave computations along with quantifiable strength of magnetic exchange interactions. Furthermore, magnetic-field induced enhanced spin-phonon coupling corroborates the proposition of magnetic-ion lattice displacement being the dominant factor for the ME behavior in this family of systems. Our results emphasize that the powder-averaged THz absorption spectrum acquired on a polycrystalline sample is not a limitation, rather an advantage over single crystals to facilitate faster screening of magnetic materials for spin-wave excitation mode, thus, expediting the search for potential materials for THz magnonic applications.

**Supporting Information**

Supporting Information is available from the Wiley Online Library or from the author.

**Acknowledgements**

D.S.R. thanks the Science and Engineering Research Board (SERB), Department of Science and Technology, New Delhi, for financial support under research Project No. CRG/2020/002338. K.S. thanks SERB for financial support under research Project No. CRG/2021/007075. B.S.M thanks Prime Minister Research Fellowship (PMRF; 0401968) funding agency, Ministry of Education, New Delhi, and Dr. Sunil Nair for providing $Mn_4Ta_2O_9$ sample.

**Conflict of Interest**

Authors declare no conflict of interest.

**Data Availability Statement**

The data that support the findings of this study are available from the corresponding author upon reasonable request.

# SUPPORTING INFORMATION

# Myriad of Terahertz Magnons with All-Optical Magnetoelectric Functionality for Efficient Spin-Wave Computing in Honeycomb Magnet $Co_4Ta_2O_9$

*Brijesh Singh Mehra[1], Sanjeev Kumar[1], Gaurav Dubey[1], Ayyappan Shyam[1], Ankit Kumar[1], K Anirudh[1], Kiran Singh[2], Dhanvir Singh Rana[1]\**

[1]Indian Institute of Science Education and Research Bhopal, 462066, India

[2]Dr. B. R. Ambedkar National Institute of Technology, Jalandhar, 144011, India

Email: dsrana@iiserb.ac.in

## S1: Experimental Details:

### A) Sample Preparation and Magnetic Characterization:

Polycrystalline sample of $Co_4Ta_2O_9$ was prepared from solid state reaction route. The stoichiometric amount of $Co_3O_4$ and $Ta_2O_5$ (99.99% purities) powders were ground and calcined in air at 1000°C for 10 h. Sample was reground, pressed, and sintered at 1100°C for 10 h. The outcome was phase-pure disc-shaped sample (diameter ~ 7 mm & thickness ~ 600 µm). Phase purity was confirmed at room temperature by PANalytical ''Empyrean' powder X-ray diffractometer (PXRD) with Cu K$_\alpha$ radiation (1.54 Å). [Figure SI 1] Rietveld refinement analysis provided a good fit with $\chi^2 = 1.8$ and lattice parameters a=b=0.5173 nm, and c=1.415 nm.

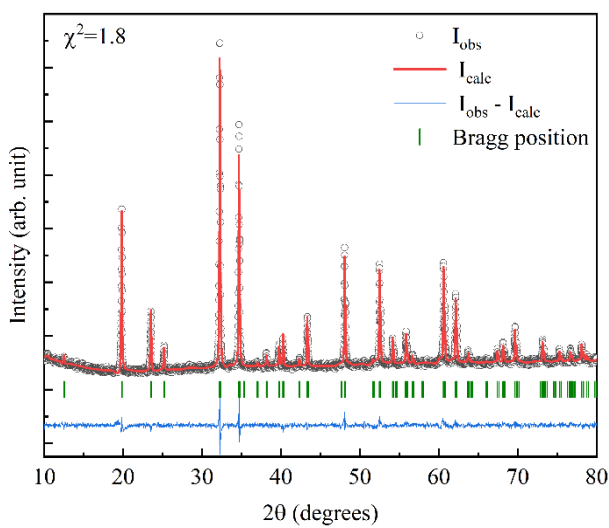

***Figure SI 1:*** *Powder X-ray diffraction of $Co_4Ta_2O_9$ depicting single-phase formation of $Co_4Ta_2O_9$.*

Magnetic measurements were performed using a superconducting quantum interference device [SQUID-VSM (Quantum Design)] in the temperature range of 2-80 K.

### B) Magneto-THz time-domain Spectroscopy:

Fiber-coupled TeraK15 THz time-domain transmission spectrometer equipped with top-loading closed-cycle He cryostat and Oxford Spectramag split-coil magnet (magnetic field up to 7T) was implemented in Faraday geometry [Figure SI 2] to measure the absorption coefficient in the spectral range 0.1-2 THz with a spectral resolution of 0.0146 THz. The path of the THz radiation is purged with nitrogen gas ten minutes before and during measurement to circumvent the water absorption peaks. THz measurement generates raw data in the form of time-dependent picosecond pulses of electric fields, which are then transformed into complex-valued frequency functions via Fast Fourier transformation. Absorption Coefficient was calculated by,

$$\alpha(\omega, T) = \frac{2}{d} \log_e \frac{S_{sample}(\omega,T)}{S_{reference}(\omega,T)},$$

where d is the thickness of the sample, $S_{sample}$ and $S_{reference}$ are the spectral amplitude with and without the sample, respectively. THz study on the polycrystalline sample provides an averaged-THz spectrum permitting us to observe excitations over all the spatial directions.

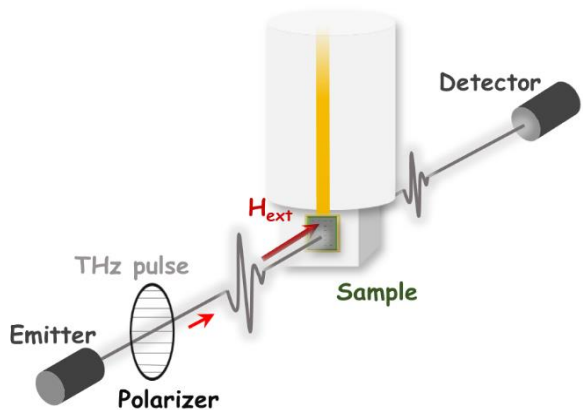

*Figure SI 2: Experimental set up of magneto-THz time-domain spectroscopy in Faraday geometry.*

**S2: Magneto-THz time-domain Experiment:**

**A) $Co_4Ta_2O_9$**

In $Co_4Ta_2O_9$, there is one magnetic-field induced excitation termed as gapless/Goldstone mode in the absorption spectrum. At 0 T it is not present but as the magnetic-field is increased it linearly blue shifts and appears in our detectable spectrum. Using its frequency versus magnetic-field trace, Landé g-factor of 3.09 was obtained which suggests the presence of unquenched orbital moments in it.

**Goldstone Mode:**

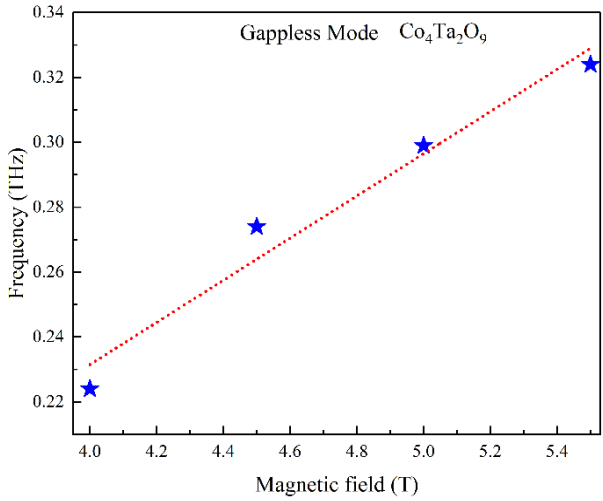

*Figure SI 3: Frequency vs magnetic-field plot for gapless excitation s7 in Co$_4$Ta$_2$O$_9$ at 6 K.*

**B) Co$_4$Ta$_2$O$_9$ and Mn$_4$Ta$_2$O$_9$**

Co$_4$Ta$_2$O$_9$ and Mn$_4$Ta$_2$O$_9$ are isostructural members of A$_4$B$_2$O$_9$ family. In the magnetic-ordered state spins of Mn$^{2+}$ in Mn$_4$Ta$_2$O$_9$ have uniaxial anisotropy spins along the c-direction whereas Co$^{2+}$ in Co$_4$Ta$_2$O$_9$ possess strong basal plane anisotropy [Figure SI 3 inset (a)]. Figure SI3 shows THz spectra with and without the sample at 10 K which emphasizes the presence (/absence) of the gapped modes in Co$_4$Ta$_2$O$_9$ (/Mn$_4$Ta$_2$O$_9$).

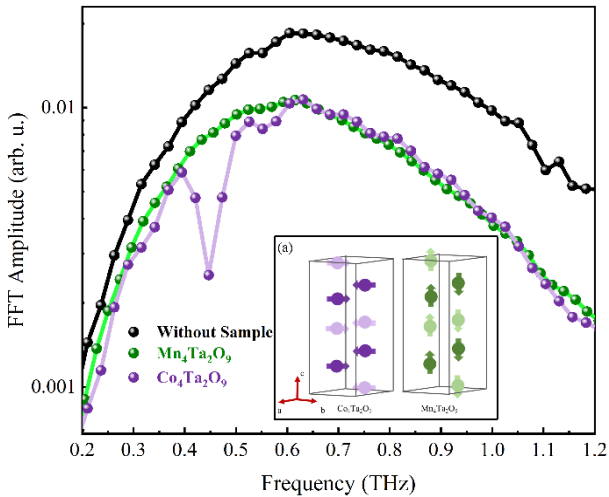

*Figure SI 4: THz spectra with (Mn$_4$Ta$_2$O$_9$ and Co$_4$Ta$_2$O$_9$) and without the samples at 10 K. Note: [Mn$_4$Ta$_2$O$_9$ sample is taken from Ref (1)]*

**C) Spin gapped and Phonon lifetime in Co$_4$Ta$_2$O$_9$:**

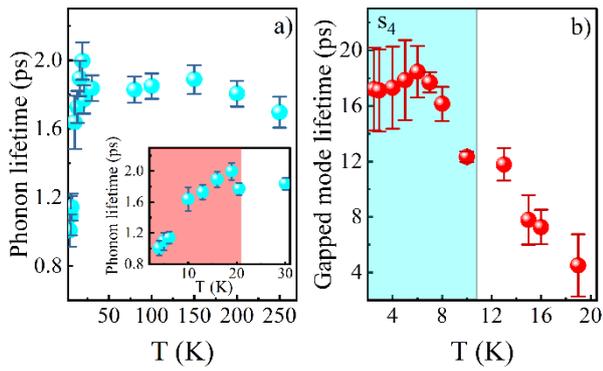

*Figure SI 5:* a) Phonon lifetime and b) Gapped mode $s_4$ lifetime as a function of Temperature.

## D) $Co_4Nb_2O_9$

**Goldstone Mode:**

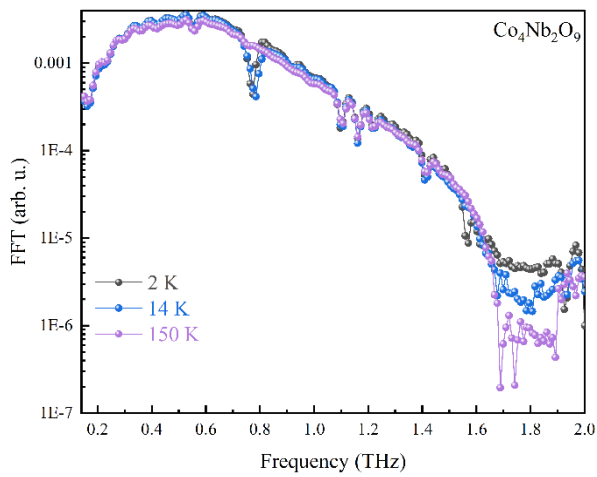

*Figure SI 6:* FFT THz spectra of $Co_4Nb_2O_9$ which shows gapped excitations and a spin-phonon coupled vibration which becomes pure phonon vibration above $T_N \sim 28$ K.

Electric field in frequency domain shows the presence of gapped mode and spin-phonon coupled mode for $Co_4Nb_2O_9$ at 0.77 THz (~3.15 meV) and 1.6-2 THz (~6.6-8.2 meV).

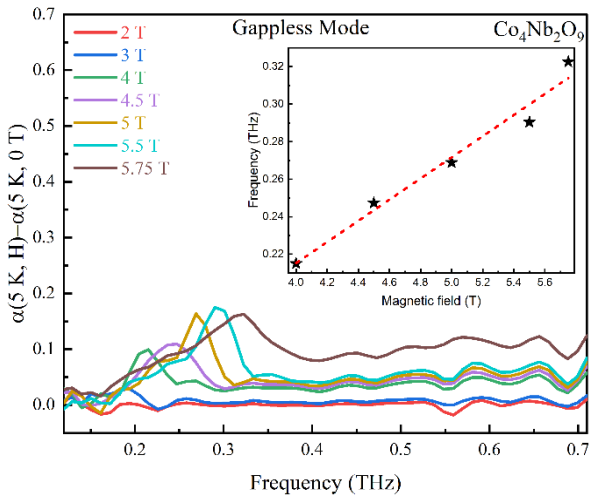

*Figure SI 7: Frequency vs magnetic-field plot for gapless excitation in $Co_4Nb_2O_9$ at 5 K.*

In $Co_4Nb_2O_9$, like $Co_4Ta_2O_9$, at 0 T the gapless excitation is not present but as the magnetic-field is increased it linearly blue shifts and appears in our detectable spectrum. Using its frequency versus magnetic-field trace, Landé g-factor of 2.68 was obtained which suggests the presence of unquenched orbital moments in it as well.

**E) Magneto-THz time-domain experiment: $Co_4Ta_{2-x}Nb_xO_9$ (x=0.3, 1, 1.7, 2)**

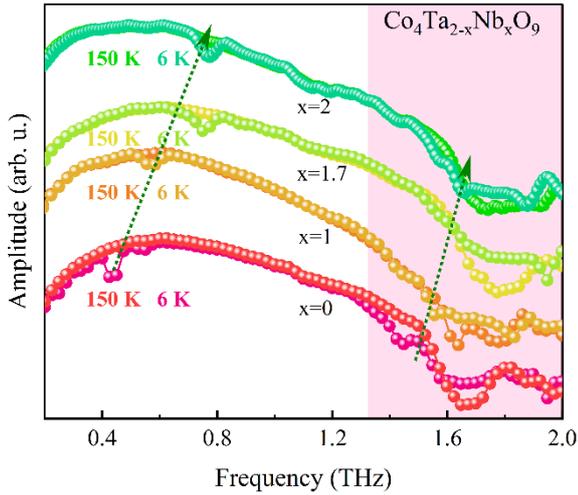

*Figure SI 8: Zero magnetic-field THz spectra for $Co_4Ta_{2-x}Nb_xO_9$ at various x values (x=0, 1, 1.7, 2). Offset is provided in the plot for clarity.*

**References:**

[1] S. N. Panja, P. Manuel, S. Nair, *Phys. Rev. B* **2021**, *103*, 014422.